\documentclass[12pt,twoside]{report}
\usepackage{indentfirst}
\newif\ifpdf
\ifx\pdfoutput\undefined
\pdffalse 
\else
\pdfoutput=1 
\pdftrue \fi
\usepackage[authoryear]{natbib}
\usepackage{esdiff}
\usepackage{ifthen}
\usepackage{amstext}
\usepackage{amssymb}
\usepackage{version}
\usepackage{bm}
\usepackage{type1cm}
\usepackage{euscript}
\usepackage{bbding}
\usepackage{fleqn}
\usepackage{epsfig}
\usepackage{epic,eepic}
\usepackage{float}
\usepackage{color}
\usepackage[unicode,bookmarks,bookmarksopen,bookmarksopenlevel=0,colorlinks,linkcolor=blue,citecolor=red]{hyperref}
\usepackage[fleqn]{amsmath}
\usepackage{mathrsfs}

\AtBeginDocument{%
  \setlength\abovedisplayskip{2pt}
  \setlength\belowdisplayskip{2pt}}

\usepackage{fancyhdr}
\fancypagestyle{plain}{%
\fancyhead{} 
}

\DeclareMathOperator{\const}{const}

\newcommand\vPi{\bm{\Pi}}

\newcommand\vu{\bm{u}}

\newcommand\vX{\bm{X}}

\newcommand\vx{\bm{x}}

\newcommand\veta{\bm{\eta}}

\newcommand\vxi{\bm{\xi}}

\renewcommand{\theequation}{\arabic{section}.\arabic{equation}}

\begin{document}

\DeclareGraphicsExtensions{.jpg,.pdf,.mps,.png}

{ \makeatletter
\renewcommand{\@oddhead}{\hfil\thepage\hfil}
\makeatother \thispagestyle{empty} \vspace*{-10mm}

\begin{center}

{\Large {\bf Discover the GLM and pseudo-Lagrangian equations of fluid dynamics on four pages}}
\\[3mm]
{\textsf{by V.\,A.~Vladimirov,\ University of York,\ vv500@york.ac.uk}
} 

\end{center}
}





\section*{Abstract}\label{abstract}

The General Lagrangian Mean (GLM) theory uses a set of averaged equations of fluid dynamics to describe interactions between mean flows and waves.
These equations are formulated in coordinates that follow the fluid's average velocity and are often referred to as `pseudo-Lagrangian' or `semi-Lagrangian'. 
This paper focuses on the principles for deriving the pseudo-Lagrangian and GLM equations, using an inviscid, incompressible, homogeneous fluid as a demonstration case.
Our exposition differs methodically from that of others and is aimed at the learners of the subject.

 \emph{Keywords:} fluid flows, pseudo-Lagrangian description, GLM theory, inviscid incompressible fluid, Lagrangian displacements, mean flows, waves, averaged equations.
\vspace{-9mm}

\section{Introduction}

The General Lagrangian Mean (GLM) theory was introduced in 1978 by Andrews and McIntyre [1] to investigate wave-mean flow interaction in geophysical contexts, following earlier attempts by Bretherton [3].
A similar theory was developed in magnetohydrodynamics by Soward in 1972 [8, 7] in the context of a pseudo-Lagrangian interpretation of the 'nearly axisymmetric geodynamo theory' by Braginskii [2, 7]. 
The approaches [1] and [8] study different classes of fluid motions.
These pioneering papers are difficult to follow in detail, whereas reading the monographs [4, 5] requires significant effort.
Recent journal publications are advanced; therefore, we do not review them.
The purpose of this paper is to clarify the mathematical origin of the GLM equations 
by exposing the pseudo-Lagrangian description of fluid flows, thereby making the theory more accessible to a broad readership.
The distribution of materials is as follows.
Section 2 explains the meaning of the pseudo-Lagrangian description.
Section 3 presents the Euler equations for an inviscid, incompressible, and homogeneous fluid in their pseudo-Lagrangian form.
Section 4 indicates how this form produces the averaged GLM equations.
Section 5 presents ideas for solving these equations.
Because of the introductory character of the paper, we avoid any discussion.
\vspace{-3mm}

\section{What is the `pseudo-Lagrangian' description?}\label{sect2}
\setcounter{equation}{0}
\vspace{-2mm}
Let us introduce two smooth and invertible vector-functions $\vx$ and $\hat\vx$ in the three-dimensional space $\vX=(X_1,X_2,X_3)$ and time $t$ as follows:
\begin{eqnarray}
&&\vx=\vx(\vX,t),\qquad  \vX=\vX(\vx,t),\label{Lagr-Eul}\\
&&\hat\vx=\hat\vx(\vX,t),\qquad  \vX=\vX(\hat\vx,t).\label{Lagr-Marker}
\end{eqnarray}
\vspace{-0mm}
Eqn.\eqref{Lagr-Eul} describes a fluid movement through the link between the Lagrangian $\vX$ and Eulerian $\vx$ coordinates of the material particles, 
and \eqref{Lagr-Marker} represents the motion of auxiliary markers that can be chosen arbitrarily, as convenient to a researcher.
In the following, we call the former \textit{an actual motion}, and the latter \textit{a reference motion}.
Two one-to-one mappings $\vx\Leftrightarrow\vX,\ \hat\vx\Leftrightarrow\vX$ yield, by excluding $\vX$, the one-to-one correspondence between the actual and reference motions:
\begin{equation}\label{strange}
    \vx\Leftrightarrow\hat\vx, \quad\text{or}\quad \ \vx=\vx(\hat\vx,t),\ \hat\vx=\hat\vx(\vx,t).
\end{equation}
As a result, any function that describes the motion of a fluid can be expressed through any of the three independent variables $(\vX,t)$, $(\vx,t)$, or $(\hat\vx,t)$.
The description of fluid motion $\vx(\hat\vx,t)$ is called \textit{pseudo-Lagrangian}, in contrast to the Lagrangian description $\vx(\vX,t)$.
The two types of velocities are
\begin{equation}
{\vu}\equiv \partial{\vx}/\partial t|_{\vX},\qquad \hat{\vu}\equiv\partial\hat{\vx}/\partial t|_{\vX},\label{velocities}
\end{equation}
where the first is \textit{the actual velocity} and the second is \textit{the reference velocity}; the subscripts after the bars denote the variables kept constant.
There are three different expressions for the same `material' derivative,
\begin{eqnarray}\label{mat-der}
&& {\partial}/{\partial t}|_{\vX}=
({\partial}/{\partial t}|_{\vx}+{\vu}\cdot\nabla)=({\partial}/{\partial t}|_{\hat\vx}+{\hat\vu}\cdot\hat\nabla)\ \ \text{or}\ \ D_{\vX}= D=\hat D.
\end{eqnarray}
This symbolic equality of three operators means that changing the independent variables should automatically entail the corresponding forms of the operators.

As an example of the transformation of the governing equation of motion to a pseudo-Lagrangian form, we use an inviscid, incompressible, and homogeneous fluid,
\begin{equation}
D u_i=-{\partial p}/{\partial  x_i},\quad {\partial u_k}/{\partial x_k}=0,\quad  D\equiv {\partial}/{\partial t}+{u_k}\,{\partial u_i}/{\partial x_k},
   \ \ i=1,2,3,\dots,\label{EulerEqn}
  \end{equation}
where $\vx=({x}_1,{x}_2,{x}_3)$ are the actual Cartesian coordinates, $\vu=(u_1,u_2,u_3)$  is the actual velocity field, $p= p(\vx,t)$ is the pressure divided by the constant density.
Then, the pseudo-Lagrangian form of \eqref{EulerEqn} is obtained by changing the independent variables $(\vx,t)$ to $(\hat\vx,t)$,
\vspace{-3mm}
\begin{eqnarray}\label{Pseudo-Eqn}
&&J_{ik}\hat D u_i=-p_{,k},\qquad J^{-1}_{ki}\partial u_i/\partial \hat x_k =0;\\ 
&&\hat D\equiv \partial/\partial t+\hat u_k\partial/\partial \hat x_k,\quad
    J_{ik}\equiv\partial x_i/\partial \hat x_k,\quad J_{ki}^{-1}\equiv \partial \hat x_k/\partial x_i, \nonumber
\end{eqnarray}
where the summation convention over repeating subscripts is introduced.
The unknown functions in \eqref{Pseudo-Eqn} are $\vx(\hat\vx,t)$, $\vu(\hat\vx,t)\equiv\vu(\vx(\hat\vx,t),t)$, $p(\hat\vx,t)$, and $\hat\vu(\hat\vx,t)$.
The latter represents the reference motion introduced by an arbitrary function $\hat\vx(\vX,t)$ in \eqref{Lagr-Marker}-\eqref{velocities}.
Mathematically, \eqref{Pseudo-Eqn} can be solved after this function is given by a researcher.
 
\vspace{-7mm}
\section{Two faces of Lagrangian displacements}
\vspace{-2mm}
The inconvenience of Eqs.\eqref{Pseudo-Eqn} is its too general form, which is not adapted to the study of perturbations of main flows and wave motions.
Such an adaptation is the split of both functions \eqref{strange} into two addends,
\begin{equation}\label{split}
 \vx(\hat\vx,t)\equiv\hat\vx+\vxi(\hat\vx,t),\qquad \hat\vx(\vx,t)\equiv\vx+\veta(\vx,t),
\end{equation}
where the roles of $\vxi$ and $\veta$ are `to measure' the deviation of the actual positions of fluid particles from the positions of markers or, mathematically, 
the deviation of the mapping  $\vx\Leftrightarrow\hat\vx$ from the identity $\vx=\hat\vx$.
The independent variables are
\begin{equation}\label{split-A}
\vxi=\vxi(\hat\vx,t),\qquad \veta=\veta(\vx,t)=\veta(\hat\vx+\vxi(\hat\vx,t),t)\equiv\veta^\xi.
\end{equation}
An immediate outcome of \eqref{split} is:
\begin{eqnarray}\label{split-1}
\veta(\vx, t)=-\vxi(\hat\vx,t),\quad\text{or}\quad  \veta(\hat\vx+\vxi(\hat\vx,t), t)=-\vxi(\hat\vx,t),\quad\text{or}\quad \veta^\xi=-\vxi,
\end{eqnarray}
where we introduce the notation [4] for any function $\phi$:
\begin{eqnarray}
    \phi\equiv\phi(\hat\vx,t),\qquad\phi^\xi\equiv\phi(\hat\vx+\vxi(\hat\vx,t),t)=\phi(\vx,t).
\end{eqnarray}
In the following, $\vxi$ is used as a primary function, while $\veta$ is expressed in terms of $\vxi$.
Applying \eqref{mat-der} to \eqref{split} yields
\begin{equation}\label{mat-der1}
\hat D\vxi=-(D\veta)^\xi=\vu^\xi -\hat\vu,\quad \text{or}\quad \vu(\vx,t)=\vu^\xi=\hat\vu+\hat D\vxi.
\end{equation}
Its use along with \eqref{strange}-\eqref{mat-der},\eqref{split} allows one to transform \eqref{EulerEqn},\eqref{Pseudo-Eqn} into the form (the related calculations are bulky and omitted)
\begin{eqnarray} 
&&\hat D^2\xi_i+\xi_{k,i}\hat D \hat u_k+ \hat D (\hat u_i-\Pi_i)+\hat u_{k,i}(\hat u_k-\Pi_k)=- p^{*}_{,i},\label{Euler-full}\\
&&\hat u_{i,i}=-\hat D \xi_{i,i}-\eta_{ki}\hat D \xi_{i,k}; \label{cont-A}\\
&&{\Pi}_i\equiv-{\xi}_{k,i} \hat D\xi_k,\ p^*\equiv  p^\xi-  u_k^\xi u_k^\xi/2,\nonumber\\
&&\eta_{ki}(\vx,t)\equiv\partial \eta_k(\vx,t)/\partial x_i=(\partial \eta_k^\xi /\partial \bar x_m)(\partial\bar x_m /\partial x_i)=
\nonumber\\
&& -\xi_{k,m}(\partial\bar x_m /\partial x_i)=- \xi_{k,m}J_{mi}^{-1},\ J_{mi}^{-1}\equiv \{\delta_{mi}+\xi_{m,i}\}^{-1}, \nonumber
\end{eqnarray}
where for the transformations of $\eta_{ki}(\vx,t)$ we used $\veta^\xi=-\vxi$ \eqref{split-1}.
The shortcuts  $(\cdot )_{,i}\equiv\partial/\partial \hat x_i$ and $(\cdot )_{,t}\equiv\partial/\partial t$ denote the partial derivatives; however, to avoid confusion, this notation is not used for the partial derivatives with respect to $\vx$ and $\vX$
Eqs.\eqref{Euler-full},\eqref{cont-A} represent a system of exact governing equations of fluid dynamics written in the independent variables $(\hat\vx,t)$ for seven unknown functions $\hat\vu(\hat\vx,t)$, $\vxi(\hat\vx,t)$, and $p^*(\hat\vx,t)$. 
This system is underdetermined, as it contains only four equations for seven unknown functions.
This is natural, since six new unknown functions $\hat\vu$ and $\vxi$ have been introduced in \eqref{strange}-\eqref{mat-der}, \eqref{split} to replace only three unknown functions $\vu(\vx,t)$ in \eqref{EulerEqn}.
The transformation presented of the governing equations \eqref{EulerEqn} to \eqref{Pseudo-Eqn} and then to \eqref{Euler-full},\eqref{cont-A} is called \textit{pseudo-Lagrangian} because, instead of considering the moving material particles $\vx=\vx(\vX,t)$ (each given by $\vX=\const$) it uses different reference points $\hat\vx=\const$, where the same moving particles are identified by $\vx=\vx(\hat\vx,t)$.
It can also be called \textit{semi-Lagrangian}, since it operates simultaneously with Lagrangian displacements $\vxi$ and Eulerian velocity $\hat\vu$.
\vspace{-8mm}

\section{Breakthrough to the GLM equations}
\setcounter{equation}{0}
\noindent The GLM theory exploits the mathematical freedom afforded by the arbitrariness of the reference motion $\hat \vx$ \eqref{Lagr-Marker} and $\hat\vu$ \eqref{velocities}.
It enforces the unique physical and mathematical meaning of $\hat\vx$ and $\hat\vu$ by assigning particular values 
\vspace{-2mm}
\begin{eqnarray}\label{assign}
    \hat\vx\mapsto\langle\vx\rangle\equiv\bar\vx,\qquad\text{and}\qquad \hat\vu\mapsto\langle\vu\rangle\equiv\bar\vu,
\end{eqnarray}
\vspace{-1mm}
where the average operation $\langle\cdot\rangle$ can be chosen differently, see [1, 4, 5].
It is important that all versions of this operation commute with all other mathematical operations in this paper.
The most popular is the ensemble averaging
\begin{eqnarray}\label{average}
\langle\vx\rangle=
\int{\vx}(\vX,t,\alpha)f(\alpha)d\alpha= \int{\vx}(\bar\vx,t,\alpha)f(\alpha)d\alpha\equiv\bar\vx,\
\int f(\alpha)\,d\alpha=1,
\end{eqnarray}
where the parameter $\alpha$ is an ensemble variable, and $f(\alpha)\ge 0$ is a distribution function. 
The two averages in \eqref{average} are equal, as the functions $\bar\vx(\vX, t)$ and $\vX(\bar\vx, t)$ do not depend on $\alpha$.
From this stage on, the first relation in \eqref{split} is replaced by
\begin{equation}\label{split-alpha}
 \vx(\bar\vx,t)=\bar\vx+\tilde\vxi(\bar\vx,t;\alpha),\qquad \langle\tilde\vxi\rangle\equiv 0,
\end{equation}
where (for each $\alpha$) the tilde stands for a narrower class of functions, compared to $\vxi$ in \eqref{Euler-full}.
The field $\bar\vx$  represents \textit{the mean motion} and $\tilde\vxi=\tilde\vxi(\bar\vx,t,\alpha)$ is \textit{the field of perturbations} or \textit{the field of Lagrangian displacements}.
Changing the notation according to \eqref{assign}-\eqref{split-alpha} and applying the average $\langle\cdot\rangle$ to \eqref{Euler-full},\eqref{cont-A}
makes all linear in $\tilde\vxi$ terms disappear and leads to the GLM equations,
\begin{eqnarray}
&&\bar{D}(\bar{u}_i-\bar{\Pi}_i) +\bar{u}_{k,i}(\bar{u}_k-\bar{\Pi}_k) = -\langle{p}^{*}_{,i}\rangle \label{Euler=GLM}\\  
&&  \bar u_{i,i}=-\langle(\bar D\tilde\xi_{i,k}) (\eta_{ki})^\xi \rangle= \langle(\bar D\tilde\xi_{i,k})  \tilde\xi_{k,m}\{\delta_{mi}+\tilde\xi_{m,i}\}^{-1}\rangle;\label{cont-av}\\
&&\bar{\Pi}_i\equiv-\langle\tilde{\xi}_{k,i} \bar D\tilde\xi_k\rangle\equiv  \bar U_i+\bar A_{ik}\bar u_k;\quad \bar{U}_i\equiv-\langle\tilde{\xi}_{m,i}\tilde{\xi}_{m,t}\rangle,\ \bar{A}_{ik}\equiv -\langle\tilde{\xi}_{m,i}\tilde{\xi}_{m,k}\rangle,\nonumber
\end{eqnarray}
where $\bar\vPi=(\bar\Pi_1,\bar\Pi_2,\bar\Pi_3)$ is called the \textit{pseudomomentum vector}.
According to \eqref{mat-der}, these equations do not contain Reynolds stresses, which represent the main difficulty in turbulence models.
Equations \eqref{Euler=GLM},\eqref{cont-av} establish physically attractive links between the mean velocity $\bar{\vu}$ and the Lagrangian displacement field $\tilde\vxi$.
However, this set of equations is also underdetermined (not closed).
\vspace{-5mm}
 
\section{How to solve the GLM equations?}

A simple way to solve \eqref{Euler=GLM},\eqref{cont-av} is to prescribe  $\tilde\vxi$ in \eqref{split-alpha}.
Then \eqref{Euler=GLM},\eqref{cont-av} represents the closed system of equations for an unknown function $\bar\vu(\bar\vx,t)$.
In the case of its growth, such a solution can be seen as a \textit{vortex dynamo}.
This formulation of the problem is similar to the \textit{kinematic dynamo} in MHD [2, 7, 8], in which the average magnetic field evolves in response to a given turbulent field.

Another possibility of upgrading \eqref{Euler=GLM},\eqref{cont-av} into a closed system is to obtain an independent evolution equation for the field $\tilde\vxi$.
The initial step in [1, 4 - 6] is to consider small-amplitude waves and weak mean flow,
\begin{eqnarray}\label{weak}
     \tilde\vxi=O(\varepsilon),\quad p^{*}=O(\varepsilon),\quad \bar\vu=O(\varepsilon^2)\quad\text{with}\quad \varepsilon\ll 1.
\end{eqnarray}
In this case, the linear evolution equations for $\tilde\vxi$ of order $O(\varepsilon)$ follow from \eqref{Euler-full},\eqref{cont-A},\eqref{average} and the final system of equations is
\begin{eqnarray}
&&\tilde\xi_{i,tt}=- p^{*}_{,i},\quad \tilde\xi_{i,i}=0.\quad  - \text{equations of order}\ O(\varepsilon)\label{linearised},\\
&&\bar{D}(\bar{u}_i-\bar{\Pi}_i) +\bar{u}_{k,i}(\bar{u}_k-\bar{\Pi}_k) = -\langle{p}_{,i}^{*}\rangle,\  \bar u_{i,i}= \langle\tilde\xi_{k,i} \bar D\tilde\xi_{i,k} \rangle  
- \text{of order}\ O(\varepsilon^2),\label{quadratic}
\end{eqnarray}
which describes mean flows, linked to linear waves $\tilde\vxi$.
It seems unusual that linear perturbations can affect the mean flow, but here is a difference between the GLM theory and the linearised theory of perturbations in steady flows: the average flow is weaker than the perturbations, see \eqref{weak}.
The next step could be to develop the model \eqref{weak}-\eqref{quadratic} by applying successive approximations to \eqref{average}-\eqref{cont-av}; see [4].
Due to the simplest example \eqref{EulerEqn} of fluid dynamics chosen, Eqn.\eqref{linearised} contains only potential solutions for $\tilde\xi_{i,tt}$, as any flow (in the chosen approximation) does not support linear wave-type motions.
Several examples of potential solutions for $\tilde\vxi$ can include combinations of point sources, vortices, or moving boundaries, which are beyond the scope of this paper, see [1, 4].
A richer example of GLM equations for a density-stratified fluid that supports internal waves is given by Grimshaw [6].

The final stage of the GLM procedure is to return the solutions $\bar\vu(\bar\vx,t)$ and $\tilde\vxi(\bar\vx,t)$ to the actual (physical) independent variables $(\vx,t)$ and the related unknown functions \eqref{EulerEqn}, which can be performed using \eqref{split}-\eqref{mat-der1}.
This operation can be physically important, for example, when modifying wave shapes.
However, for the approximation \eqref{weak}-\eqref{quadratic}, one can use $\bar\vx\simeq\vx$ and $\veta^\xi\simeq-\tilde\vxi$.
For the variety of GLM equations and their solutions, see [1, 4 - 6, 8].

\vspace{2mm}
This paper originated from the author's discussion with Professor A.D.D.Craik in 2009.
In the same year, it was communicated to Professor R.H.J.Grimshaw.
The author acknowledges Professor H.K. Moffatt and Professor O. Bühler for recent discussions.
The text was edited with AI.

\vskip 3mm
{\textbf{References:}
\vskip 1mm
1. Andrews, D., McIntyre, M.E. 1978, An exact theory of nonlinear waves on a Lagrangian-mean flow.
{\it J. Fluid Mech.}, \textbf{89}, 609-646.
\vskip 1mm

2. Braginskii, S.I., 1964. Self-excitation of a magnetic field during the motion of a highly-conducting fluid. \textit{JETP}, \textbf{20}, 726-735.
\vskip 1mm

3. Bretherton, F.P., 1971. The general linearised theory of wave propagation, \textit{Lectures Appl. Math.}, \textbf{13}, 61-102.
\vskip 1mm

4. Bühler, O. 2009. \textit{Waves and mean flows.} CUP.
\vskip 1mm

5. Craik, A.D.D. 1985, \emph{Wave interactions and fluid flows.} CUP.
\vskip 1mm

6. Grimshaw, R. 1979. Mean flows induced by internal gravity wave packets propagating in a shear flow. \textit{Phil. Trans.  Royal Society of London A,} \textbf{292}(1393), 391-417.

7. Moffatt, K. and Dormy, E., 2019. \textit{Self-exciting fluid dynamos.} CUP.


8. Soward, A.M., 1972. A kinematic theory of large magnetic Reynolds number dynamos. \textit{Phil. Trans. Royal Society of London A,} \textbf{272}(1227), 431-462.


\end{document}